\documentclass[12pt]{article}
\usepackage{a4,amsmath,amssymb,cite,graphicx,ytableau}
\usepackage{rotating}
\newtheorem{theorem}{Theorem}[section]
\newtheorem{mydef}[theorem]{Definition}
\newtheorem{prop}[theorem]{Proposition}
\newtheorem{conj}[theorem]{Conjecture}

 \def\BN{\mathbb{N}}

\catcode`@=11 \@addtoreset{equation}{section} \catcode`@=12

\begin{document}
\bibliographystyle{utphys}
\begin{titlepage}
\renewcommand{\thefootnote}{\fnsymbol{footnote}}
\noindent
{\tt IITM/PH/TH/2011/5}\hfill
{\tt arXiv:1105.6231} \\[4pt]
\mbox{}\hfill 
\hfill{\textbf{v2.0}}

\begin{center}
\large{\sf  On the asymptotics of higher-dimensional partitions}
\end{center} 
\bigskip 
\begin{center}
{\sf Srivatsan Balakrishnan, Suresh Govindarajan\footnote{suresh@physics.iitm.ac.in} and  Naveen S. Prabhakar} \\[3pt]
\textit{Department of Physics, Indian Institute of Technology Madras,\\ Chennai 600036, India \\[4pt]
}
\end{center}
\bigskip
\bigskip
\begin{abstract}
We conjecture that the asymptotic behavior  of the numbers of solid
(three-dimensional) partitions is  identical to  the asymptotics of the 
three-dimensional MacMahon numbers. Evidence is provided by an exact enumeration
of solid partitions of all integers $\leq 68$ whose numbers are reproduced with
surprising accuracy using the asymptotic formula (with one free parameter) and
better accuracy on increasing the number of free parameters. We also conjecture
that similar behavior holds for higher-dimensional partitions and provide some
preliminary evidence for four and five-dimensional partitions.
\end{abstract}
\end{titlepage}
\setcounter{footnote}{0}
\begin{center}
\textit{The purpose of computation is insight, not numbers.} -- Richard Hamming
\end{center}
\section{Introduction}

Partitions of integers appear in large number of areas such number theory,
combinatorics,  statistical physics and string theory. Several properties of
partitions, in particular, its asymptotics (the Hardy-Ramanujan-Rademacher
formula) can be derived due to its connection with the Dedekind eta function
which is a modular form\cite{Hardy:1917,Rademacher:1937}. In 1916, MacMahon
introduced higher-dimensional partitions as a natural generalization of the
usual partitions of integers\cite{MacMahon}. He also conjectured generating
functions for these partitions and was able to prove that his generating
function for plane (two-dimensional) partitions was the correct one. However it
turned out that his generating function for dimensions greater than two turned
out to be incorrect. Even for plane partitions, one no longer has nice modular
properties for the generating function. Nevertheless, the existence of a
generating function enables one to derive asymptotic formulae for the numbers of
plane partitions\cite{Wright:1931}. The inability to do the same with
higher-dimensional partitions (for dimensions $>2$) has meant that these 
objects have not been studied extensively. The last detailed study, to the best
of our knowledge, is due to Atkin et. al.\cite{Atkin:1967}.

Higher-dimensional partitions do appear in several areas of physics (as well  as
mathematics) and thus it is indeed of interest to understand them better. It is
known that the  infinite state Potts model  in $(d+1)$ dimensions gets related
to $d$-dimensional partitions\cite{Wu:1997,Huang:1997}. They also appear in the
study of directed compact lattice animals\cite{Bhatia:1997}; in the counting of
BPS states in string theory and supersymmetric field
theory\cite{Feng:2007ur,Lucietti:2008cv}. For instance, it is known that the
numbers of mesonic and baryonic gauge invariant operators in some
$\mathcal{N}=1$ supersymmetric field theories get mapped to higher-dimensional
partitions\cite{Feng:2007ur}. The Gopakumar-Vafa (Donaldson-Thomas) invariants
(in particular, the zero-brane contributions) are also related to deformed
versions of higher-dimensional partitions (usually plane
partitions)\cite{Gopakumar:1998ii,Gopakumar:1998jq}(see also
\cite{Behrend:2009}). 

In this paper, we address the issue of asymptotics of higher-dimensional 
partitions as well as explicit enumeration of higher-dimensional
partitions. Our work builds on the seminal work of Mustonen and Rajesh on the
asymptotics of solid partitions\cite{Mustonen:2003}. lack of a simple
formula for the generating functions of these partitions has been a significant
hurdle in their study. The conjectures on the asymptotics of higher dimensional
partitions given in this paper, even if partly true, would constitute progress
in the study of higher-dimensional partitions. The conjecture on the asymptotics
was arrived upon serendipitously by us when we found that a one-parameter
formula for solid-partitions derived using MacMahon's generating function worked
a lot better than it should. To be precise, a formula that was meant to obtain
an order of magnitude estimate (for solid partitions of integers in the range
$[50,62]$) was not only getting the right order of magnitude  but was also
correct to $0.1-0.5\%$ (around 3-4 digits). The main conjecture discussed in
section 3 is a natural outgrowth of this observation. The exact enumeration of
solid partitions was possible due to an observation that lead to a gain of the
order of $10^4$ to $10^5$ enabling us to exactly generate numbers of the order
of $10^{16}-10^{17}$ in reasonable time. 

The paper is organized as follows. Following the introductory section,  section
2 provides the background to problem of interest as well as fixes the notation.
Section 3 deals with asymptotics of higher-dimensional partitions. This done by
means of two conjectures. We provide some evidence towards these conjectures
with a fairly detailed study of solid partitions using a combination of exact
enumeration as well as fits to the data. Section 4 provides the theoretical
background to the method used for the exact enumeration of higher-dimensional
partitions. We conclude in section 5 with some remarks on extensions of this
work. In appendices A we work out the asymptotics of MacMahon numbers. Appendix
B provides an `exact' asymptotic formula for three-dimensional MacMahon numbers.
In appendix C we present several tables that includes our results from exact
enumeration as well some details of the fits for solid partitions.

\section{Background}
A partition of an integer $ n $, is a weakly decreasing sequence $ (a_0,a_1,a_2,\ldots ) $ such that  
\begin{itemize}
\item $ \sum_i a_i= n $ and 
\item $ a_{i+1}\leq a_{i}\quad \forall\ i $. 
\end{itemize}
For instance, $ (2,1,1) $ is a partition of $ 4 $.  Define $ p_1(n) $ to be the
number of partitions of $ n $. For instance,
\begin{equation}
4=3+1=2+2=2+1+1=1+1+1+1 \quad \implies \quad p_1(4)=5\ .
\end{equation}
A slightly more formal way  definition of  a partition is as  a map from  $
\mathbb{Z}_{\geq0} $ to $ \mathbb{Z}_{\geq0} $  satisfying the two conditions
mentioned above. This definition enables one to generalise to higher dimensional
partitions. A \textit{$d$-dimensional partition} of $ n $ is defined to be a map
from $ \mathbb{Z}_{\geq0}^d $ to $ \mathbb{Z}_{\geq0} $ such that it is weakly
decreasing along all directions and the sum of all its entries add to $ n $. Let
us denote the partition by $ (a_{i_1,i_2,\ldots,i_d}) $. The
weakly decreasing
condition along the $ r$-th direction implies that 
\begin{equation}
a_{i_1,i_2,\ldots,i_r+1,\ldots, i_d}\leq a_{i_1,i_2,\ldots,i_r,\ldots i_d}\quad 
\forall\  (i_1,i_2,\ldots,i_d)\ .
\end{equation}
Two-dimensional partitions are also called \textit{plane} partitions while
three-dimensional  partitions are also called \textit{solid} partitions. Plane
partitions can thus be written out as a two-dimensional array of numbers,
$a_{ij}$. For instance, the two-dimensional partitions of $4$ are
\begin{multline}
\begin{smallmatrix} 4 \end{smallmatrix} \quad 
\begin{smallmatrix} 3 & 1 \end{smallmatrix} \quad 
\begin{smallmatrix} 3 \\ 1 \end{smallmatrix} \quad
\begin{smallmatrix} 2 & 2 \end{smallmatrix} \quad
\begin{smallmatrix} 2 \\ 2 \end{smallmatrix} \quad
\begin{smallmatrix} 2 &  1 & 1\end{smallmatrix} \quad
\begin{smallmatrix} 2 & 1 \\ 1  \end{smallmatrix} \quad
\begin{smallmatrix} 2 \\ 1 \\ 1  \end{smallmatrix} \quad
\begin{smallmatrix} 1 & 1 & 1 &1 \end{smallmatrix} \quad
\begin{smallmatrix} 1 & 1 & 1 \\ 1 \end{smallmatrix} \quad
\begin{smallmatrix} 1 & 1 \\ 1 &1 \end{smallmatrix} \quad
\begin{smallmatrix} 1 & 1 \\ 1 \\ 1 \end{smallmatrix} \quad
\begin{smallmatrix} 1 \\ 1 \\ 1 \\ 1 \end{smallmatrix} \quad
\end{multline}
Thus we see that there are $13$ two-dimensional partitions of $4$.  Let us
denote by $ p_d(n) $ the number of $d$-dimensional  partitions of $ n
$.\footnote{We caution the reader that there is another definition of
dimensionality of a partition that  differs from ours. For instance, plane
partitions would be three-dimensional partitions in the nomenclature used in
Atkin et. al.\cite{Atkin:1967} while we refer to them as two-dimensional
partitions.} It is useful to define the generating function of these partitions
by ($ p_d(0)\equiv1 $)
\begin{equation}
P_d(q) \equiv \sum_{n=0}^\infty p_d(n)\ q^n\ . \label{generatingfunction}
\end{equation}
The generating functions of one and two-dimensional partitions have very nice
product representations.  One has the \text{Euler formula} for the generating
function of partitions
\begin{equation}
P_1(q) = \frac1{\prod_{n=1}^\infty (1-q^n)}\ ,
\end{equation}
and the \textit{MacMahon formula} for the generating function of plane partitions
\begin{equation}
P_2(q) = \frac1{\prod_{n=1}^\infty (1-q^n)^n}\ .
\end{equation}
MacMahon also guessed a product formula for the generating functions  for $d>2$
that turned out  to be wrong\cite{Atkin:1967}. His guess is of the form
\begin{equation}
M_d(q)= \frac1{\prod_{n=1}^\infty (1-q^n)^{\binom{n+d-2}{d-1}}}
:=\sum_{n=0}^\infty m_d(n)\ q^n\ .\label{MacMahonformula}
\end{equation}
We will refer to the numbers $m_d(n)$ as  the $d$-dimensional MacMahon numbers.
It  is easy to see that $M_1(q)=P_1(q)$ and $M_2(q)=P_2(q)$. However $M_d(q)\neq
P_d(q)$ for $ d>2 $. An explicit  formula (given by Atkin et.
al.\cite{Atkin:1967} or  the book by Andrews\cite{AndrewsPartitions}) for the
number of $ d $-dimensional partitions of $ 6 $ is 
\begin{equation}
p_d(6)=1+ 10d + 27 \binom{d}2 + 28 \binom{d}3 + 11\binom{d}4 + \binom{d}5\ .
\end{equation}
Then, one can show that 
\begin{equation}
m_d(6)-p_d(6)=\binom{d}3+\binom{d}4\ ,
\end{equation}
which is non-vanishing for $ d\geq 3 $. Thus the MacMahon generating function
fails to  generate numbers of partitions when $ d\geq3$.

\subsection{Presentations of higher-dimensional partitions}

There are several ways to depict higher dimensional partitions. Recall that
there is a one  to one correspondence between (one-dimensional) partitions of
$n$ and Ferrers (or Young) diagrams. The partition of $4$ corresponding to $3+1$
corresponds to the Ferrers diagram 
$$
\ytableausetup{smalltableaux}\ydiagram{3,1}\ .
$$
Similarly, the plane partition $\begin{smallmatrix} 3 \\ 1 \end{smallmatrix}$
can  be represented by a Young tableau (i.e., a Ferrers diagram with numbers in
the boxes) or as a `pile of cubes' stacked in three dimensions (one of the
corners of the cubes being located at $(0,0,0)$, $(0,0,1)$, $(0,0,2)$ and
$(1,0,0)$ in a suitably chosen coordinate system)
\begin{center}
\ytableausetup{aligntableaux=bottom,nosmalltableaux}\begin{ytableau}
3 \\ 1
\end{ytableau}\qquad \qquad
\includegraphics[height=14mm]{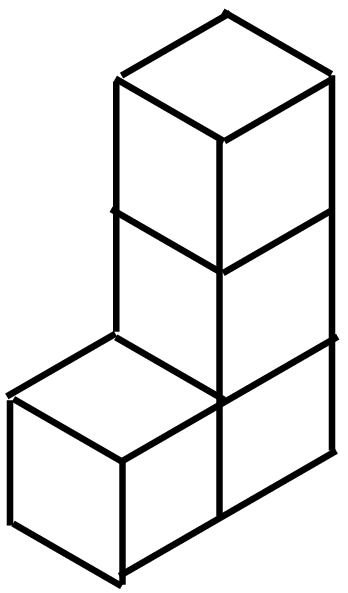}
\end{center}
Similarly, $d$-dimensional partitions can be represented as a pile of hypercubes in $(d+1)$ dimensions.

We refer the reader to  the work by Stanley (and references therein) for an
introduction  to plane partitions\cite{Stanley:1971,Stanley:1985}. The book by
Andrews\cite{AndrewsPartitions} provides a nice introduction to
higher-dimensional partitions. Further the lectures by Wilf on integer
partitions\cite{Wilf:2000} and the notes by Finch on partitions\cite{Finch:2004}
are also good starting points to existing literature on the subject.

\section{Asymptotics of higher-dimensional partitions}

In this section, we will discuss the asymptotics of higher-dimensional
partitions.  The absence of an explicit formula for the  generating function for
$d>2$ implies that there is no simple way to obtain the asymptotics of such
partitions. In this regard, an important result due to Bhatia et. al. states
that\cite{Bhatia:1997}
\begin{equation}
\lim_{n\rightarrow\infty} n^{-d/d+1} \log p_d(n) = \textrm{$d$-dependent constant.}
\end{equation}

\begin{conj}\label{weakconj} The constant in the above formula is identical  to
the one for the corresponding MacMahon numbers.
\begin{equation}
\lim_{n\rightarrow\infty} n^{-\frac{d}{d+1}} \log p_d(n) = 
\lim_{n\rightarrow\infty} n^{-\frac{d}{d+1}} \log m_d(n) = \frac{d + 1}d \Big[
d\ \zeta(d + 1)\Big]^{\frac1{d + 1}}=: \beta_1^{(d)}\ .
\end{equation}
\end{conj}
For three-dimensional partitions, this becomes  a conjecture of  Mustonen and
Rajesh. Mustonen and Rajesh used  Monte-Carlo simulations to compute the
constant  and showed that it is $1.79\pm0.01$\cite{Mustonen:2003}. This is
compatible with the conjecture since $\beta_1^{(3)}\sim 1.78982$.

It is important to know the sub-leading behavior of the asymptotics of
higher-dimensional  partitions in order to have   quantitative estimate of
errors. This is something  we will provide in the next subsection. Before
discussing the asymptotic behavior of the higher-dimensional partitions, it is
useful to know the asymptotic  behavior of the MacMahon numbers. A calculation
shown in appendix A gives their sub-leading behavior. One obtains
\begin{equation}\label{MacMahonasymptotics}
\log m_d(n) \sim  \sum_{r=1}^{d}\  \beta^{(d)}_r\ n^{\frac{d-r+1}{d+1}} + \gamma^{(d)} \log n + \delta^{(d)} \ .
\end{equation}
The constants $\beta^{(d)}_r$ and $\gamma^{(d)}$ have been computed  for $d=3,4,5$ in appendix A.

\subsection{Towards a stronger conjecture}

The number of $d$-dimensional partitions of $n$ can be obtained from the
generating function $P_d(q)$  by inverting Eq. \eqref{generatingfunction}
\begin{equation}
p_d(n) = \frac1{2\pi}\int_{-\pi}^\pi P_d(e^{iy})\ e^{-iny}\ dy\ . \label{theintegral}
\end{equation}
Suppose we knew all the singularities of the function $P_d(q)$. The integral can
be then be evaluated  (at large $n$), for instance, by the saddle point method
and adding up the contribution of all singularities thus obtaining an asymptotic
formula for $p_d(n)$.  The singularities  are usually obtained by looking at product formulae of the form
\begin{equation}
P_d(q) = \prod_{n=1}^\infty (1-q^n)^{-a^{(d)}(n)}\ .
\end{equation}
The exponents $a^{(d)}(n)$ can be determined for those values of $n$ for which 
$p_d(n)$ has been determined.  If all the $a^{(d)}(n)$ are positive, then it is
easy to see that $P_d(q)$ is singular at all roots of unity -- this leads
naturally to the circle method of Hardy and Ramanujan\cite{Hardy:1917}. However,
for $d>2$, this turns out to be false. For instance, $a^{(3)}(15)=-186$ is the
first exponent that becomes negative for $d=3$\cite[see Table 1]{Knuth:1970}. We
will assume that the singularities of $P_d(q)$ continues to occur at roots of
unity. In particular, we will see that the Bhatia et. al. result implies that
for large enough $n$, one has
\begin{equation}
a^{(d)}(n) = \mathcal{O}(n^{d-1})\ ,
\end{equation}
with $a^{(d)}(n)>0$. Let us assume that the dominant term in a saddle point
computation of  the integral in Eq. \eqref{theintegral} occurs near $q=1$. 
\begin{prop}\label{LaurentExpansion} The Laurent expansion of  $\log
P_d(e^{-t})$ in  the neighbourhood of $t=0$ is of the form
\begin{equation}
-\log P_d(e^{-t}) = \frac{\widehat{C}_d}{d\, t^d} +
\frac{\widehat{C}_{d-1}}{(d-1)\,t^{d-1}}  + \cdots +\frac{\widehat{C}_1}{t} +
\textrm{non-singular as } t \rightarrow 0\ ,
\end{equation}
where $\widehat{C}_1,\ldots,\widehat{C}_d$ are some constants.
\end{prop}
\textbf{Remark:} This is precisely the form of the Laurent expansion for   $\log
M_{d}(e^{-t})$ near $t=0$ (see Appendix A).

A saddle point computation of the integral \eqref{theintegral}  is carried out  
by extremizing the function
$$
\log P_d(e^{-t}) + n t \ .
$$
The extremum, $t_*$, which is close to $t=0$ for large $n$,   obtained using
Proposition \ref{LaurentExpansion} is given by
\begin{equation}
t_* = \left(\frac{\widehat{C}_{d}}{n}\right)^{1/(d+1)} +  \cdots 
\end{equation}
Plugging in the saddle point value, we see that 
\begin{align}
\log p_d(n) &\sim   \frac{\widehat{C}_d}{d\,t_*^d} +
\frac{\widehat{C}_{d-1}}{(d-1)\,t_*^{d-1}}  + \cdots +\frac{\widehat{C}_1}{t_*}
+ n t_* + \cdots \\
& \sim \frac{d+1}d \left(\widehat{C}_{d}\right)^{1/(d+1)}n^{d/(d+1)}+ \textrm{sub-leading terms}\ .
\end{align}
We thus recover the  bound obtained by Bhatia et. al.\cite{Bhatia:1997}.  Thus,
we see  that the Bhatia et. al. result combined with the assumption that 
$P_d(e^{-t})$ is a meromorphic function in the neighborhood of $t=0$ with a pole
of order $d$ implies Proposition \ref{LaurentExpansion}.

A more precise saddle point computation enables us to determine sub-leading terms as well and we obtain 
\begin{align}
\log p_d(n)\sim  \sum_{r=1}^{d}\  \widehat{\beta}^{(d)}_r\ n^{\frac{d-r+1}{d+1}}
+\widehat{\gamma}^{(d)}  \log n + \widehat{\delta}^{(d)} + \cdots  \ ,
\label{saddlepoint}
\end{align}
where the constants $\widehat{\beta}^{(d)}_r$, $\widehat{\gamma}^{(d)}$ and
$\widehat{\delta}^{(d)}$ are  determined  by the constants $\widehat{C}_r$ that
appear in Proposition \ref{LaurentExpansion}.

Conjecture \ref{weakconj}  implies that $\widehat{C}_d=d\, \zeta(d+1)$ -- this
is the leading coefficient in  the Laurent expansion of $\log M_{d}(e^{-t})$
near $t=0$. This is equivalent to 
\begin{equation}
a^{(d)}(n) = \frac{n^{d-1}}{(d-1)!}+ \cdots\ ,
\end{equation}
where the ellipsis indicates sub-leading terms in the large $n$ limit.  We  now
propose a stronger form of conjecture \ref{weakconj}.
\begin{conj}\label{strongconj} The asymptotics of the $d$-dimensional partitions
are identical  to the asymptotics of the MacMahon numbers.
\begin{equation}\label{asymptoticformula}
\log p_d(n) \sim  \sum_{r=1}^{d}\  \beta^{(d)}_r\ n^{\frac{d-r+1}{d+1}} + \gamma^{(d)} \log n + \cdots  \ ,
\end{equation}
where $\beta^{(d)}_r$ and $\gamma^{(d)}$ are as in Eq. \eqref{MacMahonasymptotics}.
\end{conj}
It is easy to see that one can have conjectures that are stronger than
conjecture \ref{weakconj}  but weaker than conjecture \ref{strongconj} by
requiring fewer coefficients to match with Eq. \eqref{MacMahonasymptotics}. 
Conjecture \ref{strongconj} implies that the coefficients, $\widehat{C}_r$
($r=1,\ldots,d$)  in the Laurent expansion in Proposition \ref{LaurentExpansion}
are identical to those of $\log M_d(e^{-t})$. Equivalently, 
\begin{equation}
P_d(e^{-t})-M_d(e^{-t}) = \mathcal{O}(1)\ ,
\end{equation}
near $t=0$. It also implies that at large $n$, $a^{(d)}(n)$ behaves exactly like
the exponent  that appears in the product formula for $d$-dimensional MacMahon
numbers in Eq. \eqref{MacMahonformula}, i.e.,
\begin{equation}
a^{(d)}(n) \sim \binom{n+d-2}{d-1}+ \cdots \ ,
\end{equation}
where the ellipsis indicates terms that vanish as $n\rightarrow\infty$.

\subsection{Evidence for the conjecture}

We will provide evidence by explicitly enumerating numbers for the
higher-dimensional partitions.  In particular, we compute all solid partitions
for $n\leq 68$ and use the formula provided by Eq. \eqref{asymptoticformula} as
a one-parameter function to fit known numbers. The advantage of this procedure
is that one doesn't need to go to enormously large values of $n$. In Figures 1,
2 and 3, we compare this formula implied by conjecture \ref{strongconj} for $d=3,4,5$
respectively. Since the values of $n$ that we consider are not too large, these
fits provide weak evidence that three of the conjectured numbers i.e., 
$\beta^{(d)}_1$, $\beta^{(d)}_2$ and $\gamma^{(d)}$ are probably correct.
\begin{table}
\centering
\begin{tabular}{crr} \hline
$N$ & $q_3(N)$\hspace{1cm} & $p_3(N)$\hspace{1cm} \\[3pt] \hline
 58 & \underline{397}2318521718539 & 3971409682633930 \\
 59 & \underline{652}2014363273781 & 6520649543912193 \\
 60 & \underline{1068}6367929548727 & 10684614225715559 \\
 61 & \underline{1747}4590403967699 & 17472947006257293 \\
 62 & used to fit constant & 28518691093388854 \\ 
 63 & \underline{4645}3074905306481&46458506464748807 \\
 64 & \underline{755}22726337662733 & 75542021868032878 \\
 65 & \underline{122}556018966297693 & 122606799866017598 \\
 66 & \underline{198}518226269824763 & 198635761249922839 \\
 67 & \underline{32}0988410810838956 & 321241075686259326 \\
 68 & \underline{518}102330350099210 &518619444932991189 \\[3pt] \hline
\end{tabular}
\caption{Estimates using the asymptotic formula $q_3(N)$. The constant in the 
asymptotic formula is fixed by requiring it to give the exact answer for $N=62$
-- the largest known number of solid partitions at the time of the fit. }
\end{table}

\begin{figure}[ht]
\centering
\includegraphics[height=2.3in]{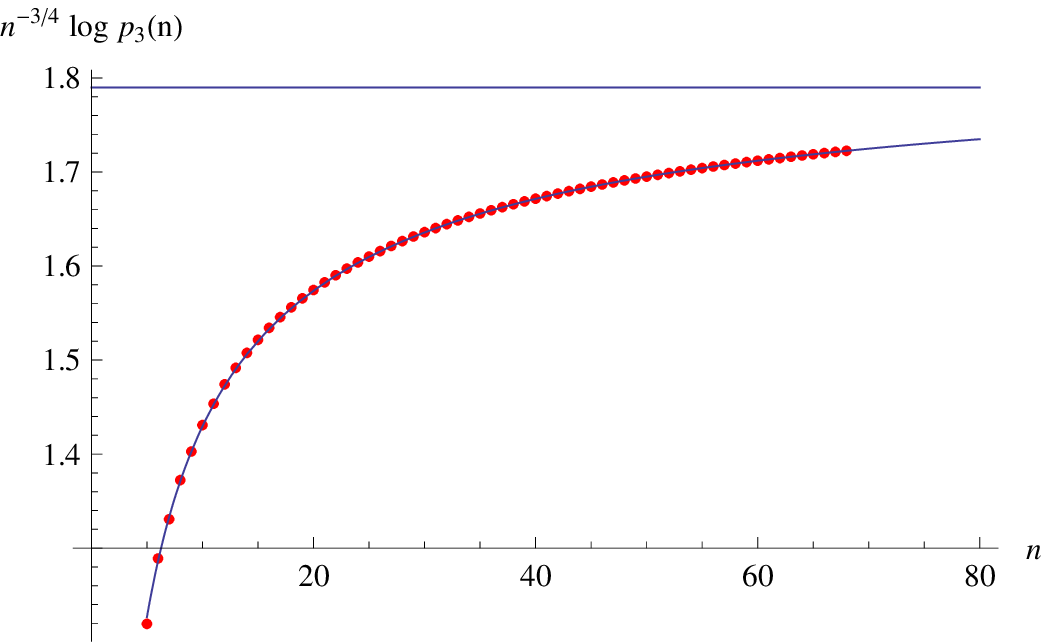}
\caption{Plot of $n^{-3/4}\log p_3(n)$ for $n\in [5,68]$ (red dots). The blue
curve is  the asymptotic formula normalized to give the correct answer for
$n=62$ and the horizontal line is the conjectured value for $n\rightarrow
\infty$.}
\end{figure}

\begin{figure}[ht]
\centering
\includegraphics[height=2.5in]{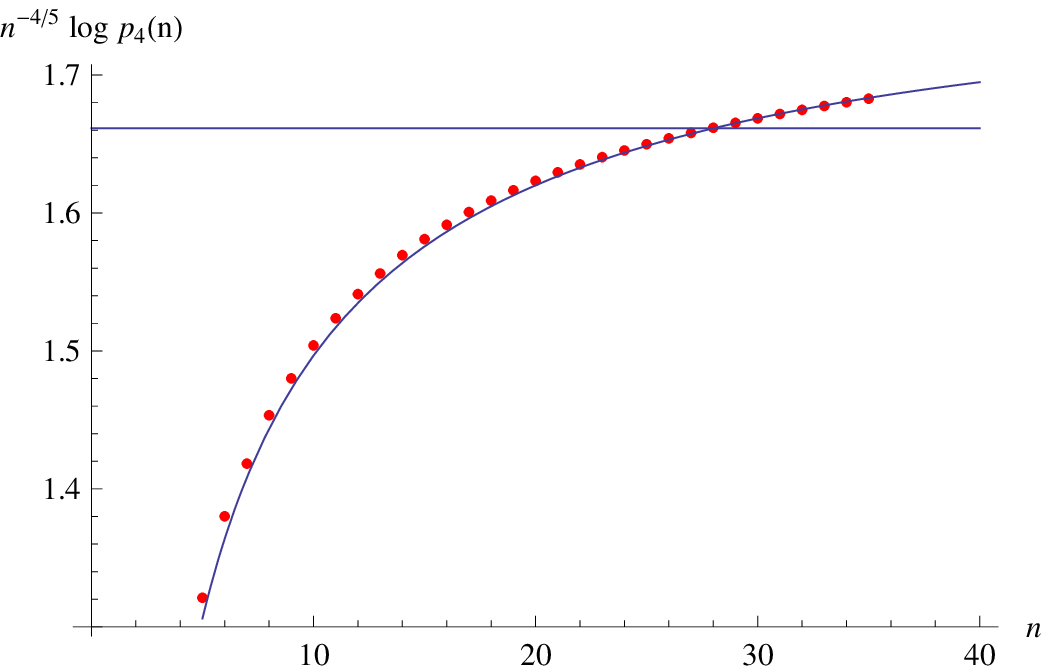}
\caption{Plot of $n^{-4/5}\log p_4(n)$ for $n\in [5,35]$ (red dots).  The blue
curve is the asymptotic formula normalized to give the correct answer for $n=30$
and the horizontal line is the conjectured value for $n\rightarrow \infty$.}
\end{figure}

\begin{figure}[ht]
\centering
\includegraphics[height=2.5in]{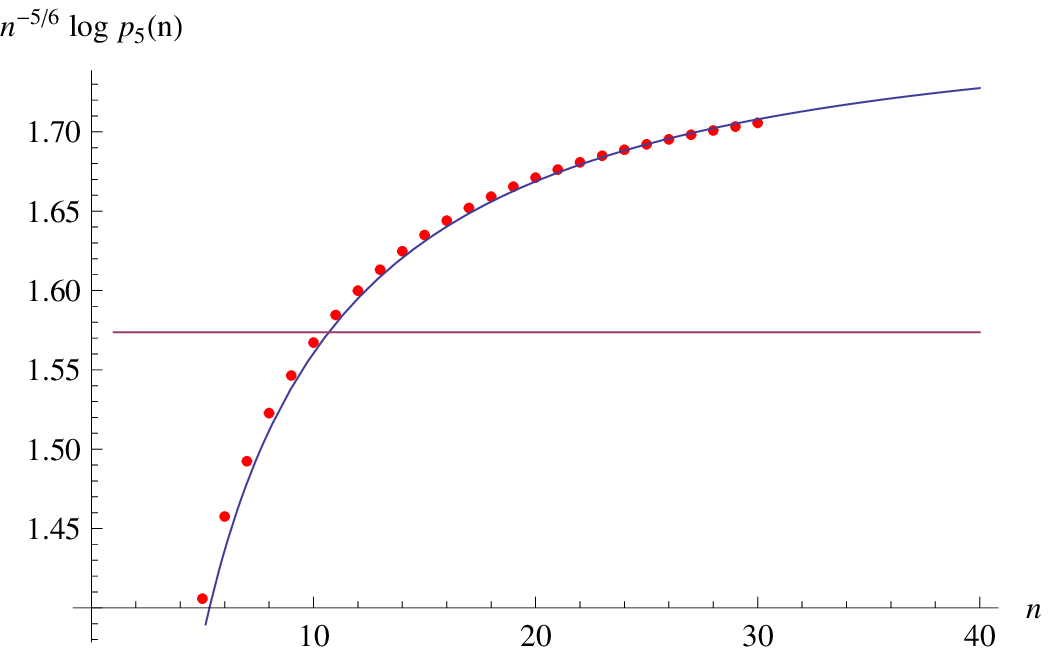}
\caption{Plot of $n^{-5/6}\log p_5(n)$ for $n\in [5,30]$ (red dots). The blue
curve is the asymptotic  formula normalized to give the correct answer for
$n=25$ and the horizontal line is the conjectured value for $n\rightarrow
\infty$.}
\end{figure}

\subsection{Solid partitions: a detailed study}

The asymptotic expansion of the logarithm of three-dimensional MacMahon numbers
is (with  $\xi\equiv n+\tfrac{\zeta(-3)}4$)
\begin{equation}
\log m_3(n) \sim \frac43 [3 \zeta(4)]^{1/4}\ \xi^{3/4} + 
\frac{\zeta(3)}{2[3\zeta(4)]^{1/2}}\ \xi^{1/2}  -\frac{\zeta(3))^2}{8
[3\zeta(4)]^{5/4}}\ \xi^{1/4} - \frac{61}{96} \log \xi + \cdots 
\end{equation}
Using the above formula as a guide, we fit the solid partitions to the following
 three formulae involving up to three parameters $(a,b,c)$: ($\xi:=n+b$)
\begin{align*}
q_3(n)&=  \frac43 [3 \zeta(4)]^{1/4}\ n^{3/4} + 
\frac{\zeta(3)}{2[3\zeta(4)]^{1/2}}\ n^{1/2}  -\frac{\zeta(3))^2}{8
[3\zeta(4)]^{5/4}}\ n^{1/4} - \frac{61}{96} \log n + a \\
r_3(n)&=  \frac43 [3 \zeta(4)]^{1/4}\ \xi^{3/4} + 
\frac{\zeta(3)}{2[3\zeta(4)]^{1/2}}\ \xi^{1/2}  -\frac{\zeta(3))^2}{8
[3\zeta(4)]^{5/4}}\ \xi^{1/4} - \frac{61}{96} \log \xi + a \\
s_3(n)&=  \frac43 [3 \zeta(4)]^{1/4}\ \xi^{3/4} + 
\frac{\zeta(3)}{2[3\zeta(4)]^{1/2}} \ \xi^{1/2} -c\ \xi^{1/4} - \frac{61}{96}
\log \xi + a \ .
\end{align*}
Note that the number of free parameters increases from $1$ for the function $q_3$ to $2$ for $r_3$ and  to $3$ for
$s_3$. 
We obtain $a=-1.544$,
$(a,b)=(-1.530,-0.028)$ and $(a,b,c)=(-3.211,1.689,0.257)$ from the three fits.
We use the same functions to estimate the values of three-dimensional MacMahon
numbers  for the same range of values using a similar fit.
We see that the function $s_3$ has worked almost as well as it did for
the corresponding MacMahon numbers. In particular, the fit gives
$c=0.25713$ which is different from the one given by MacMahon numbers for $
\beta^{(3)}_3=-0.041413$.  For the MacMahon numbers, the fitted value of
$c=-0.057621$ which is close to the actual number.
This suggests that the coefficient of $n^{1/4}$ may be different from the one
given by  the MacMahon numbers.  For the values of $n$ that we have considered, the
dominant contributions are due to the first two terms as well as the log term.
Hence, we consider this as possible evidence for $\widehat{\beta}^{(3)}_r =
\beta^{(3)}_r$ for $r=1,2$.  For completeness, we provide the numbers obtained by carrying out a five-parameter fit using the numbers in the range $[60,68]$. The fit gives:
\begin{equation}
\log p_3(n) \sim 1.73\ n^{3/4} + 0.83\ n^{1/2} -0.90\ n^{1/4} - 1.00\ \log n -0.22\ .
\end{equation}
We also observe that if we used a larger range of numbers, say, $n\in [50,68]$, we obtain  large numbers (of order ten or greater) for some of the coefficients. This reflects the lack of data for large number more than anything else. 


In an attempt at understanding the accuracy of our numbers better,  we carried
out a systematic study of an exact asymptotic formula (in the sense of
Hardy-Ramanujan-Rademacher for partitions)  for three-dimensional MacMahon
numbers using a method due to Almkvist\cite{Almkvist:1993,Almkvist:1998}. These
are discussed in Appendix B. One writes
\[
m_3(n) \sim \sum_{k=1}^\infty \phi_k(n) \ ,
\]
where $\phi_k(n)$ are the contributions from various saddle-points with $k=1$ being the dominant one. For $n=60$, we see that $\phi_1(60)$ gets the first nine digits right while the sum of the first two terms get eleven digits right. We further broke up the contribution of $\phi_1(n)$ into several terms. The term that we write as $\phi_1^{(0)}(n)$ is the contribution from the singular part of $\log M_3(e^-t)$ at the dominant saddle point located near $t=0$. We see that $\phi_1^{(0)}(60)$ gets the first five digits right -- somewhat closer to what we have obtained in our estimates for the numbers of solid partitions. 
 
\subsection{An unbiased estimate for the leading coeffficient}

\begin{figure}[ht]
\centering
\includegraphics{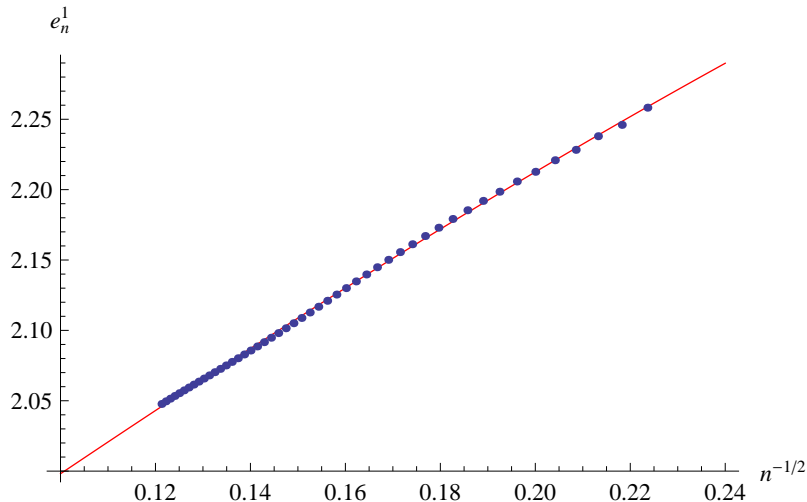}
\caption{A plot of $e_n^1$ vs $n^{-1/2}$ for $n\in[20,68]$ along with a
three-parameter fit.}\label{eonefig}
\end{figure}

\begin{figure}[hbt]
\centering
\includegraphics{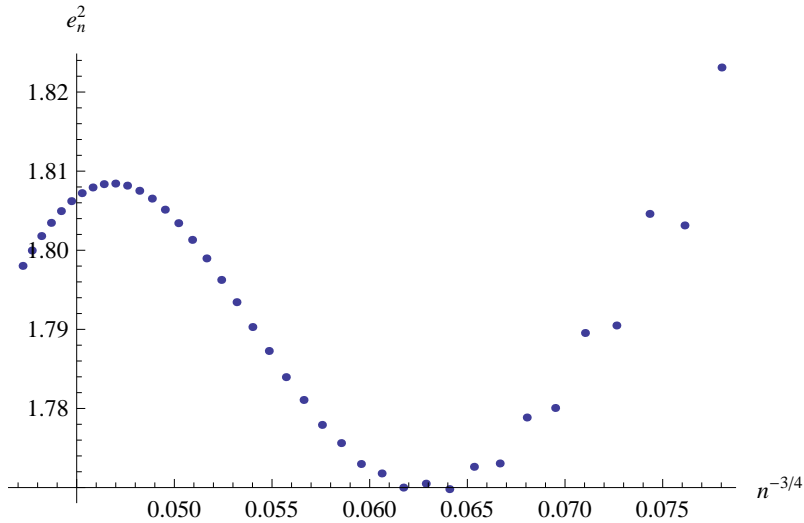}
\caption{A plot of $e_n^2$ vs $n^{-3/4}$ for $n\in[20,68]$.}\label{etwofig}
\end{figure}

In order to provide an unbiased estimate for the \textit{leading} coefficient of
the asymptotic formula using the exact numbers of solid partitions\footnote{We
thank the anonymous referee for suggesting that we provide an unbiased estimate
of the leading coefficient and for asking us to look at the methods discussed in
ref. \cite{GG}. }, we use the method of Neville tables (albeit with a slight and
obvious modification)\cite{GG}. Let
\begin{align}
e_n^0 &\equiv n^{-3/4}\log p_3(n) \nonumber \\
&\sim \sum_{x=1}^3 \hat{\beta}_x^{(3)} n^{(1-r)/4} + \hat{\gamma}^{(3)}
n^{-3/4}\log n + \hat{\delta}^{(3)} n^{-3/4}\ ,
\end{align}
where we have written the asymptotic formula in the second line using the
parameters defined in Eq. \eqref{saddlepoint}. Further, for $r\geq1$,
recursively define 
\begin{equation}
e_n^r := \frac{n^{1/4}\ e_n^{r-1}-(n-r)^{1/4}\
e_{n-1}^{r-1}}{n^{1/4}-(n-r)^{1/4}}\ .
\end{equation}
Using the conjectured asymptotic formula for $p_3(n)$, we can derive asymptotic
formulae for $e_n^r$. The $e_n^r$ have been constructed so that 
\begin{enumerate}
\item $\lim_{n\rightarrow \infty} e_n^r$ tends to a constant that equals
$\hat{\beta}^{(3)}_1$ for all $r$. The first sub-leading term is proportional to
$n^{-(r+1)/4}$. Thus a plot of $e_n^r$ vs $n^{-(r+1)/4}$ should be a straight
line in the asymptotic limit.
\item As we increase $r$, the number of parameters that appear in the asymptotic
formula for $e_n^r$ decrease. For instance, one sees that $\hat{\beta}_2^{(3)}$
drops out for $r=1$:
\begin{equation}
e_n^1 \sim \hat{\beta}^{(3)}_1 - \hat{\beta}_3^{(3)} n^{-1/2} -2
\hat{\gamma}^{(3)} n^{-3/4}\log n + (4\hat{\gamma}^{(3)}-2 \hat{\delta}^{(3)})
n^{-3/4}\ , \label{eonepar}
\end{equation}
and $\hat{\beta}_2^{(3)},\ \hat{\beta}_3^{(3)}$ drop out for $r=2$:
\begin{equation}
e_n^2 \sim \hat{\beta}^{(3)}_1+ \hat{\gamma}^{(3)} n^{-3/4}\log n +
(-6\hat{\gamma}^{(3)}+ \hat{\delta}^{(3)}) n^{-3/4}\ ,
\end{equation}
\end{enumerate}
An estimate for $\hat{\beta}^{(3)}_1$ has been obtained by carrying out two and
three-parameter fits to the asymptotic formula given in Eq. \eqref{eonepar}. We
obtain
\begin{equation}
e_n^1 =\left\{\begin{matrix} 1.793+2.099 n^{-1/2} & \textrm{two-parameter fit}\\
1.781 + 0.83 n^{-1/2} + 0.924 \log n  & \textrm{three-parameter fit}
\end{matrix}
\right.
\end{equation}
A four-parameter fit leads to coefficients that are not of order one. We discard
this fit as we make the natural assumption that all coefficients are of order
one or smaller. Using the two different fits, we can estimate
$\hat{\beta}^{(3)}_1$ is around $1.78-1.79$. The wide variation that we observe
in $\hat{\beta}^{(3)}_2$ suggests that we cannot estimate it with the available
exact numbers. In figure \ref{eonefig}, we have plotted $e_n^1$ vs $n^{-1/2}$
along with the three-parameter fit. We also observe that $e_n^2$ (see figure
\ref{etwofig}) is oscillating between $[1.77,1.81]$ and hence we cannot estimate
any further parameters using the data. For completness, we have carried out a
similar analysis for the MacMahon numbers, $m_3(n)$ in the range $n\in [20,68]$
and obtain $\beta^{(3)}_1$ in the range $[1.77-1.78]$. We also observe that
$e_n^2$ does not oscillate as it does for solid partitions. 

We conclude that an
unbiased estimate for $\hat{\beta}^{(3)}$ is consistent with conjecture  \ref{weakconj}.
However, given the relatively small values of $n$ that we have used, this only
constitutes weak evidence at best. There is another result due to Widom et. al.
who studied the asymptotics of (restricted) solid partitions with Ferrers
diagrams that fit in a four-dimensional box of size $10^3\times p$\cite{Widom:2002} as a function of $p$.
They observe that the entropy in the thermodynamic limit deviates\footnote{The entropy for fixed boundary conditions was found to be $0.145$ instead of the conjectured value of $0.139$. See Eq. (14) in ref. \cite{Widom:2002}.} from a
formula derived from a MacMahon formula for restricted solid partitions. Should
we expect  a similar behavior for unrestricted solid partitions? The deviation
observed by Widom et. al. is small.  If a similar behavior
occurs for unrestricted partitions, then conjecture  \ref{weakconj} would be false. We believe that the exact numbers that we have used are not large enough to definitively test conjecture  \ref{weakconj}. 
However, in any case, it is important to note that that the functional form of the asympotics continues to hold. 


\section{Explicit Enumeration}

In this section, we discuss the explicit enumeration of higher dimensional partitions.
The first program to explicitly enumerate higher-dimensional partitions is due
to  Bratley and McKay\cite{Bratley:1967a}. However, we do not use their
algorithm but another one due to Knuth\cite{Knuth:1970}. We start with a few
mathematical preliminaries in order to understand the Knuth algorithm  as well
as our parallelization of the algorithm.  
 
\subsection{Almost Topological Sequences}
Let $ P $ be a set with a partial ordering (given by a relation denoted by  $
\prec $) and a well-ordering (given by a relation denoted by $ < $). Further,
let the partial ordering be embedded in the well-ordering i..e, $ x\prec y $
implies $ x<y $.
\begin{mydef}
A sequence $\mathbf{X}= (x_1,x_2,\ldots,x_m) $ containing elements of $ P $ is
called a  \textit{topological sequence} if\cite{Knuth:1970}
\begin{enumerate} 
\item For $ 1\leq j \leq  m $ and $ x \in P $, $ x \prec x_j $ implies $ x = x_i $ for some $ i < j $;
\item If $ m > 0 $, there exists $ x \in P $ such that $ x < x_m $ and $ x \neq x_i $, for $ 1 < i \leq m $.
\end{enumerate}
\end{mydef}
Let us call a $j$-th position  in a topological sequence, $ \mathbf{X} $,
\textit{interesting}   if  $ x_j> x_{j+1} $. By definition, the last position of
a sequence is considered interesting.  The index of a topological sequence is
defined to be the sum of all $ j $ for all interesting positions i.e.,
\begin{equation}
\textrm{index}(\mathbf{X}) =\sum_j \left\{j ~|~ j \textrm{ is interesting}\right\}\ .
\end{equation}
\begin{mydef}
An almost topological sequence is a  sequence that satisfies condition 1 but not necessarily condition  2.
\end{mydef}
Thus all topological sequences are also almost topological sequences. This
definition is  motivated by the observation  that almost topological sequences
do occur as sub-sequences of topological sequences. 

\subsubsection{An example due to Knuth}

Let $ P $ denote the set of three-dimensional lattice points i.e.,
\begin{equation}
P=\Big\{(i,j,k)~|~ i,j,k=0,1,2,3,\ldots \Big\}\equiv \mathbb{N}^3
\end{equation}
with the partial ordering $ (i,j,k)\preceq (i',j',k') $ if $ i\leq i' $, and $
j\leq j' $ and $ k\leq k' $.  Let us choose the well-ordering to be given by the
lexicographic ordering i.e., 
\begin{equation}
(i,j,k)< (i',j',k')
\end{equation}
if and only if
$$
i< i' \quad \textrm{ or } (i=i' \textrm{ and } j<j') \quad \textrm { or } (i=i',\ j=j' \textrm{ and } k<k')\ .
$$
The \textbf{depth} of a topological sequence is the number of elements in the
sequence.  Consider the topological sequence (of depth $6$)
$$ 
\mathbf{X}=\{(0,0,0),  (0,0,1),  (0,0,2),  \mathbf{(1,0,0)},  \mathbf{(0,1,0)},  \mathbf{(0,0,3)} \}
$$
 where we have indicated the interesting positions in boldface. This sequence has index $ 15=4+5+6 $.

\subsection{Topological sequences and solid partitions}

Let $ d_m(n) $ denote the number of topological sequences  of the set
$P=\mathbb{N}^m$ with index $ n $.  Further, define $d_m(0)=1$. As before, let
$p_m(n)$ denote the number of $m$-dimensional partitions of $ n $. A theorem of
Knuth relates these two sets of numbers as follows:
\begin{theorem}[Knuth\cite{Knuth:1970}]
\begin{equation}
p_m(n) = \sum_{k=0}^n d_m(k)\ p_1(n-k)\ .
\end{equation}
\end{theorem}
Equivalently, the generating function of $m$-dimensional partitions decomposes
into a product of  the generating function of the numbers of topological
sequences and the generating function of one-dimensional partitions. 
\begin{equation}
P_m(q) = D_m(q) \  P_1(q)\ ,
\end{equation}
where 
$$
D_m(q):=\sum_{n=0}^\infty d_m(n)\ q^n\ .
$$
Since topological sequences are much easier to enumerate, Knuth went ahead and
wrote a program to  generate all topological sequences of index $ \leq N $ (for
some fixed $ N $). This is the program that was the starting point of our exact
enumeration.

We list below the topological sequences of index 2  and 3 when $P=\BN^3$ (we
have dropped  the comma between numbers to reduce the length of the expression)
\begin{align*}
\textbf{Index 2:}&\  \big\{(000) (010)\big \} \quad \textrm{and}\quad \big\{(000) (100)\big \} \quad \implies \boxed{d_3(2)=2}\ . \\
\textbf{Index 3:}&\  \big\{(000) (001) (010)\big \}\ ; \big\{(000) (001) (100)\big \}\ ; \big\{(000) (010) (020)\big \}\ ;\\
& \big\{(000) (010) (100)\big \}  \ ;\big\{(000) (100) (200)\big \} \quad \implies \boxed{d_3(3)=5}\ .
\end{align*}
Thus, we see that $D_3(q)=1+2q^2+5q^3+\cdots$. We also have $P_1(q) = 1 + q + 2q^2 + 3q^3 +\cdots$. Thus, we obtain
$$
P_3(q) = D_3(q)\ g_1(q) = 1 + q + 4 q^2 + 10 q^3 + \cdots
$$

\subsection{Equivalence classes of almost topological sequences}

We say that two sequences  $ \mathbf{X}= (x_1,x_2,\ldots,x_m) \sim \mathbf{Y}=
(y_1,y_2,\ldots,y_m) $ are related if  the elements of $ \mathbf{Y} $ are a
permutation of the elements of $ \mathbf{X} $. Of course, not all permutations
of an almost topological sequence lead to another almost topological sequence as
some of them violate condition 1 in the definition of a topological sequence.
However, even after imposing the restriction to permutations that lead to other
topological sequences, the relation remains an equivalence relation. As an
example consider the following three sequences in $\BN^3$:
\begin{align}\label{eqclassex}
\big\{(0,0,0),  (0,0,1),  (0,0,2),  (1,0,0)\big\}\ ,\nonumber\\
\big\{(0,0,0),  (0,0,1),  (1,0,0),  (0,0,2)\big\}\ ,\\
\big\{(0,0,0),  (1,0,0),  (0,0,1),  (0,0,2)\big\}\ . \nonumber
\end{align}
It is easy to see that these three sequences form a single equivalence class.
However,  the last two are \textit{not} topological sequences as they violate
condition 2 in the definition of a topological sequence. and hence are almost
topological sequences. We thus choose to work with equivalence classes of almost
topological sequences.
\begin{prop} The equivalence classes of  almost topological sequences of $\BN^d$
of depth $k$ is  in one to one correspondence with $(d-1)$-dimensional
partitions of $k$. We shall refer to the $(d-1)$-dimensional partition as the
\textbf{shape} of the equivalence class.
\end{prop}
The $(d-1)$-dimensional partition is obtained by placing $d$-dimensional
hypercubes (of size one) at the points  appearing the almost topological
sequence. This is nothing but the `piles of cubes' representation of a
$(d-1)$-dimensional partition. In this representation, the precise ordering of
the points in the almost topological sequence is lost and one obtains the
\textit{same} $(d-1)$-dimensional partition for any element in the same
equivalence class. Given a $(d-1)$-dimensional partition, the coordinates of the
hypercubes in the `piles of cubes' representation give the elements of the
almost topological sequence.
 For instance, the equivalence class in Eq. \eqref{eqclassex}  has as
its shape the following two-dimensional partition of $4$:
$$
\begin{ytableau}
3 \\ 1
\end{ytableau}\qquad \qquad
\includegraphics[height=14mm]{31.eps}\quad .
$$

When $P=\mathbb{N}^2$, the almost topological  sequences of $P$ are standard
Young tableaux. Given an almost topological sequence of $\mathbb{N}^2$ with
shape $\lambda$ with $n$ boxes, the standard Young tableau is obtained by
entering the position of the box in the almost topological
sequence\footnote{Recall that a Young tableau is a Ferrers diagram with boxes
filled in with numbers. A standard Young tableau has numbers from $(1,\ldots,n)$
such that the numbers in the boxes increase as one moves down a column or to the
right.}. It is easy to see that this map is a bijection.
It is an interesting and open problem to enumerate the number of almost 
topological sequences given a shape for higher-dimensions. We did this by
generating all topological partitions of a given index and sorting them out by
shape. However, this is an overkill if one is interested in enumerating 
topological sequences associated with a particular shape.

\subsection{Programming Aspects}
 
 The explicit enumeration of topological sequences to generate partitions was
first carried out Knuth who  enumerated solid partitions of integers $\leq
28$\cite{Knuth:1970}. This was extended to all integers $\leq 50$ by Mustonen
and Rajesh (using other methods)\cite{Mustonen:2003}. We first ported Knuth's
Algol program to C++ and quickly found that it was prohibitively hard to
generate additional numbers given the fact that $p_3(50)$ is of the order of
$10^{13}$. So we decided to parallelize Knuth's program in the following way. 
 \begin{enumerate}
 \item Generate all almost topological sequences up to a depth $k$.
 \item Next, separately run each sequence (to generate the rest of tree) from
depth $(k+1)$   until all  sequences of index $N$ that contain the initial
sequence as its first $k$ terms are generated. Here it is important to note that
while we are counting the numbers of topological sequences, we need to include
\text{all} almost topological sequences since they necessarily appear as
sub-sequences of topological sequences. 
 \item An important observation is that it suffices to run one sequence for every 
 given shape since they have identical tree structure after the $(k+1)$-th node.
However,  it is crucial to note that each topological sequence in a given
equivalence class does \text{not} have the same index. This entails a bit of
book keeping where one keeps track of the different indices of all topological
sequences of identical shape. The power of this approach is best illustrated by
looking at Table \ref{shapenumbers} where we list the numbers of actual
sequences (nodes) as well the number of shapes. A naive estimate (based on the
reduction of the number of runs) shows that run times should go down by an order
of  $10^{5}-10^{6}$.
 \end{enumerate}
 \begin{table} \centering
 \begin{tabular}{ccccc}\hline
Depth    & 12 & 14 & 15 & 17 \\[3pt] \hline
Nodes   & 28680717 & 1567344549 & 12345147705 & 856212871761\\[3pt]\hline 
 Shapes & 1479 & 4167 & 6879 & 18334 \\[3pt] \hline
 \end{tabular}
\caption{Number of equivalence classes  at various depths (equal to the number
of plane partitions) for counting  topological  partitions of $\BN^3$.}
\label{shapenumbers}
\end{table}

This approach has enabled us to extend the Knuth-Mustonen-Rajesh results to all
integers $N\leq 68$. The numbers  were generated in several steps:
$N=52,55,62,68$. The results for $N\leq52$ we obtained without parallelization.
The results for $N\leq 55$ were obtained using parallelization to depth $7$ but
without using equivalence classes and required about 1500 hours of CPU time. The
results for $N\leq 62$ were done using parallelization to depth $14$ ($4167$
shapes) and took around 30000 hours of CPU time(about a month of runtime). The
last set of results for $N\leq 68$ took around $360K$ hours of runtime (spread
over five months).

We also extended the numbers for four-dimensional partitions of $N\leq 35$ and 
five-dimensional partitions of $N\leq 30$. This was done without any
parallelization. The complete results are given in  appendix \ref{exactresults}.

\section{Conclusion}

We believe that our results show that it is indeed possible to understand the
asymptotics of higher  dimensional partitions. The preliminary nature of our
results shows that a lot more can and should be done. Our results provide a
functional form to which results from Monte Carlo simulations, of the kind
carried out by Mustonen and Rajesh\cite{Mustonen:2003}, can be fitted to.
However, the errors should be better than one part in $10^3$ or $10^4$ to be
able to fix the sub-leading coefficients.  We are indeed making preliminary
studies to see whether one can achieve this.

Another avenue is to see if there are sub-classes of partitions that can be
counted i.e.,  we can provide simple expressions for their generating functions.
For instance, the analog of conjugation in usual partitions is the permutation
group, $S_{d+1}$, for $d$-dimensional partitions. Following
Stanley\cite{Stanley:1986}, we can  organise $d$-dimensional partitions based on
the subgroups of $S_{d+1}$ under which they are invariant (see also
\cite{Krattenthaler:1990}). Some of these partitions might have simple
generating functions.

One of the proofs of the MacMahon formula for the generating function of plane
partitions is due to  Bender and Knuth\cite{Bender:1972}(see also
\cite{Nijenhuis:1978}). It is done by considering a bijection between plane
partitions and matrices with non-negative entries. There is a natural
generalization of such matrices into hypermatrices -- these hypermatrices are
counted by MacMahon numbers. It would be interesting to contruct a  Bender-Knuth
type map between solid partitions and hypermatrices and study how it fails to be
a bijection. This might  explain why  the asymptotics of MacMahon numbers works so
well for higher-dimensional partitions.

\bigskip\bigskip

\noindent \textbf{Acknowledgments:} We would like to thank Arun Chaganty,
Prakash Mohan,  S. Sivaramakrishnan as well the other undergraduate students of
IIT Madras' Boltzmann group who provided a lot of inputs to the project on the
exact enumeration of solid partitions
(\texttt{http://boltzmann.wikidot.com/solid-partitions}).
We thank the High Performance Computing Environment
(\texttt{http://hpce.iitm.ac.in}) at  IIT Madras for providing us with a stable 
platform (the leo and vega superclusters) that made the explicit enumeration of
higher-dimensional partitions possible. We thank Nicolas Destainville for
drawing our attention to ref. \cite{Widom:2002}.

\appendix

\section{Asymptotics of the MacMahon numbers}

In this appendix, we work out the asymptotics of the MacMahon numbers using a
method  due to Meinardus\cite{Meinardus:1954}. A nice introduction to this
method is found in the paper by Lucietti and Rangamani\cite{Lucietti:2008cv}. 

We have seen that the generating function for $d$-dimensional MacMahon numbers is given by
\begin{equation}
M_d(q)=1 + \sum_{n=1}^{\infty}m_d(n)\, q^n =
\prod_{n=1}^{\infty}\frac{1}{(1-q^n)^{\binom{n+d-2}{d-1}}} \ .
\end{equation}
Inverting this, we obtain:
\begin{equation} \label{eq:mm1}
m_{d}(n) = \oint_{\Gamma}\frac{dq}{2{\pi}i}\frac{M_d(q)}{q^{n+1}}
\end{equation}
where  $q$ is a complex
variable and $\Gamma$ is a circle $|q|=\varepsilon<1$ traversed in the counterclockwise direction.
We shall evaluate the contour integral in (\ref{eq:mm1}) by writing $q
= e^{-t}$ and then taking the limit $t \to 0$. This corresponds to the
contribution to \eqref{eq:mm1} due to the pole at $q = 1$, which is
the dominant contribution. The poles of $M_d(q)$ occur precisely at
all roots of unity, with the sub-dominant contributions coming from other roots of unity.

We have,
\begin{equation} \label{eq:mm3}
\log M_d(e^{-t}) = -\sum_{n=1}^{\infty}a_n\log (1-e^{-t n}), \quad a_n =
\binom{n+d-2}{d-1}.
\end{equation}
We expand the logarithm inside the sum using its Taylor series and using
the Mellin representation of $e^{-x}$ i.e.,
\begin{equation}
e^{-x} = \frac{1}{2{\pi}i}\int_{\gamma -
    i\infty}^{\gamma + i\infty}\, ds\, x^{-s}\, \Gamma(s)\ , \quad \gamma > 0\ .
\end{equation}
We obtain 
\begin{equation} \label{eq:mm4}
\log M_d(e^{-t}) = \frac{1}{2{\pi}i}\int_{\gamma - i\infty}^{\gamma +
  i\infty}\, ds \,\Gamma(s)\,\zeta(s+1)\,D_d(s)\,{t}^{-s}\ ,
\end{equation}
where the Dirichlet series $D_d(s)$ defined as
\begin{displaymath}
D_d(s) = \sum_{n=1}^{\infty}\,\frac{a_n}{n^s} \ .
\end{displaymath}
The real constant $\gamma$ is chosen to lie to the right of all poles of $D_d(s)$ in the $s$-plane.
For  $d = 3$, $a_n=n(n+1)/2$ and hence the Dirichlet series is 
\begin{displaymath}
D_3(s) = \sum_{n=1}^{\infty}\,\frac{n(n+1)}{2\,n^s} =
\tfrac12 \big[\zeta(s-2)+\zeta(s-1)\big]\ .
\end{displaymath}
Hence, $D_3(s)$ has simple poles at $s = 2, 3$ with residue $1/2$ at
both poles. For general $d$, $D_d(s)$ has poles at $s = k,\, k = 2, 3, \ldots,
d$. Let us denote the residue at $s=k$ by  $A_k$.

Now, we shift the contour in (\ref{eq:mm4}) from Re$(s) = \gamma$ to
Re$(s) = -\alpha$, for $0 < \alpha < 1$. In the process, $\log M_d(q)$
receives contributions from the poles of the integrand that lie
between Re$(s) = \gamma$ and Re$(s) = -\alpha$. Hence, we get
\begin{multline} \label{eq:mm5}
\log M_d(e^{-t})  =  \sum_{k=2}^{d}A_k\,\Gamma(k)\,\zeta(k+1){t}^{-k} + D_d'(0)
-D_d(0)\log {t} \\ 
 + \frac{1}{2{\pi}i}\int_{-\alpha - i\infty}^{-\alpha +
  i\infty} ds\, \Gamma(s)\,\zeta(s+1)\,D_d(s)\,{t}^{-s}.
\end{multline}
The integral can be shown to go as $\mathcal{O}(|t|^{\alpha})$. Hence, we
get
\begin{equation} \label{eq:mm6}
M_d(e^{-t}) = \exp \Big(\sum_{k=2}^{d}A_k\,\Gamma(k)\,\zeta(k+1)\,{t}^{-k} + D_d'(0) 
-D_d(0)\log {t} \Big) \Big(1 + \mathcal{O}(|t|^{\alpha})\Big)
\end{equation}
Hence, near $q = 1$, we have
\begin{equation} \label{eq:mm7}
m_d(n) = \frac{1}{2{\pi}i}\int_{t_0-i\pi}^{t_0+i\pi}  dt \, e^{G_d(t)}.
\end{equation}
where ($t_0$ is taken to close to $0^+$)
\begin{displaymath}
G_d(t) = \sum_{k=2}^{d}\frac{C_k}{k\,{t}^k} + nt\ , \qquad C_k := A_k\,
\Gamma(k+1)\, \zeta(k+1)
\end{displaymath}

We carry out the integral (\ref{eq:mm7}) using the saddle point
method. For this, we have to first evaluate $t = t_*$ such that
$G'_d(t_*) = 0$. That is,
\begin{equation} \label{eq:mm8}
\sum_{k=2}^{d}\frac{C_k}{t_*^{k+1}} - n = 0.
\end{equation}
We next let the integration contour pass through the saddle point for
which the value of $G_d(t_*)$ is largest. This happens when $t_*$
is the largest root of \eqref{eq:mm8}. This means $t_*^{-(d+1)}
\sim n$ or equivalently, $t_* \sim n^{-1/d+1}$ and hence, $t_*
\to 0$ as $n \to \infty$. Hence, the saddle point method indeed gives
the value of $m_d(n)$ for $n \to \infty$.  

Now, we solve for $t_*$ from (\ref{eq:mm8}) which is a polynomial
equation of degree $d+1$. For $d > 3$, we do not have a general
formula for the roots of the equation. But in this case, we indeed
have a formula for the largest positive root of (\ref{eq:mm8}), due to
Lagrange:
\begin{equation}
t_* (n)= \sum_{\substack{\ell\,>\,0 \\ \ell \neq 0\textrm{ mod }(d+1)}}^{\infty}b_\ell\, n^{-\ell/(d+1)}
\end{equation}
where
\begin{displaymath}
b_\ell= \frac{1}{\ell!}\Bigg[\frac{d^{\ell-1}}{dy^{\ell-1}}\phi(y)^\ell\Bigg]_{y=0} \textrm{ with } 
\quad \phi(y) \equiv \Bigg(\sum_{k=1}^{d}C_k\,
y^{d-k}\Bigg)^{\frac{1}{d+1}}.
\end{displaymath}
Using the above formula, we can compute $t_0$ to any required order in $n$ and
then carry out  the saddle-point integration
\eqref{eq:mm8}. We finally get
\begin{equation}
m_d(n) =\sqrt{\frac{1}{2{\pi}G_d''(t_*)}} \  t_*^{-D_d(0)}\: \exp\big(G_d(t_*) +
D_d'(0)\big)\, \Big(1 + \mathcal{O}\left(t_*^{\alpha}\right)\Big)\ .
\end{equation}
Recall that the dependence on $n$ occurs implicitly, on the right hand side of
the above equation,  through the saddle-point  value $t_*(n)$.

\subsection{Three-dimensional MacMahon numbers}

The  asymptotic formula is
\begin{equation}
m_3(n) \sim  \textrm{const } n^{-61/96} \ \exp(\widehat{G}_3(n))\ ,
\end{equation}
where\footnote{We add a term corresponding to $k=1$ with coefficient $C_1$ in
Eq. \eqref{eq:mm8} so that  the saddle point computation can be carried over for
higher-dimensional partitions for which that might be the case.}
\begin{equation*}
\widehat{G}_3(n):=\frac{4}{3} {C_3}^{1/4} n^{3/4}+\frac{C_2 }{2 C_3^{2/4}}\,n^{2/4}+\frac{\left(8
   C_1 C_3-C_2^2\right) }{8 C_3^{5/4}}\,n^{1/4}
   \end{equation*}
with $C_1=0$, $C_2=\zeta(3)$ and $C_3=3 \zeta(4)$.
Numerically evaluating, we obtain
\begin{equation}
\widehat{G}_3(n)\simeq 1.78982 n^{3/4}+0.333546 \sqrt{n}-0.0414393 n^{1/4}\ .
\end{equation}
\subsection{Four-dimensional MacMahon numbers}

The asymptotic formula is
\begin{equation}
m_4(n) \sim  \textrm{const } n^{-2179/3600} \ \exp(\widehat{G}_4(n))\ ,
\end{equation}
where
\begin{equation*}
\widehat{G}_4(n):=\frac{5}{4} {C_4}^{1/5} n^{4/5}+\frac{C_3 n^{3/5}}{3 C_4^{3/5}}+\frac{\left(5
   C_2 C_4-C_3^2\right) }{10 C_4^{7/5}}\,n^{2/5}+\frac{\left(C_3^3-5 C_2 C_4
   C_3+25 C_1 C_4^2\right) }{25 C_4^{11/5}}\,n^{1/5}
\end{equation*}
with $C_1=0$, $C_2=2\zeta(3)/3$, $C_3=3 \zeta(4)$ and $C_4=4 \zeta(5)$.
Numerically evaluating, we obtain
\begin{equation}
\widehat{G}_4(n)\simeq 1.66139\ n^{4/5} + 0.460969\ n^{3/5} + 0.0829315\ n^{2/5}-0.0345152\ n^{1/5} \ .
 \end{equation}
\subsection{Five-dimensional MacMahon numbers}
The  asymptotic formula is
\begin{equation}
m_5(n) \sim  \textrm{const } n^{-563/960} \ \exp(\widehat{G}_5(n))\ ,
\end{equation}
where
\begin{align*}
\widehat{G}_5(n):=&\frac{6}{5} C_5^{1/6}\, n^{5/6}+\frac{C_4}{4 C_5^{2/3}}\, n^{4/6}+\frac{\left(4
   C_3 C_5-C_4^2\right) }{12 C_5^{3/2}}\,n^{3/6}\nonumber \\
   &+\frac{\left(2 C_4^3-9 C_3 C_5
   C_4+27 C_2 C_5^2\right) }{54 C_5^{7/3}}\,n^{2/6}\\
   &+\frac{\left(-91 C_4^4+504
   C_3 C_5 C_4^2-864 C_2 C_5^2 C_4+432 C_5^2 \left(12 C_1
   C_5-C_3^2\right)\right)}{5184 C_5^{19/6}}\ n^{1/6}\ ,
 \end{align*}
 with $C_1=0$, $C_2=\tfrac12\zeta(3)/3$, $C_3=\tfrac{11}4 \zeta(4)$, $C_4=6 \zeta(5)$ and $C_5=5 \zeta(6)$.
Numerically evaluating, we obtain
\begin{equation}
\widehat{G}_5(n)=1.5737\ n^{5/6}+0.525874\ n^{2/3}+0.15873\ \sqrt{n}+0.0223817\
   n^{1/3}-0.0263759\ n^{1/6}\ .
   \end{equation}

\section{A rather exact formula for $m_3(n)$}

We will work out the asymptotics of the three-dimensional MacMahon numbers using
methods due to  Almkvist\cite{Almkvist:1993,Almkvist:1998}. The generating
function of three-dimensional MacMahon numbers is
\begin{equation}
M_{3}(x) = \prod_{n=1}^\infty (1-x^n)^{-n(n+1)/2}=\sum_{n=0}^\infty m_{3}(n)\ x^n\ .
\end{equation}
The integrals are evaluated using the circle method due to Hardy and Ramanujan\cite{Hardy:1917}.
The coefficients $ m_{3}(n) $ are determined from the generating function by the formula
\begin{equation}
m_{3}(n) =\frac1{2\pi} \int_{-\pi}^\pi M_{3}\big(e^{iy}\big)\ e^{-iny} dy\ .
\end{equation}
Since $ M_{3}(x) $ has poles when ever $x$ is a root of unity, the dominant 
contributions occur in the neighborhood of this point. Setting $x=\exp(iy)$, we
see that the poles occur for all $ y=2\pi h/k $ with $ (h,k)=1 $  the
contribution can be evaluated by summing over contributions from such terms. One
writes
\begin{align}
m_{3}(n) &\sim \sum_{k=1}^\infty \sum_{\substack{h=1\\ (h,k)=1}}^{k-1}
\frac1{2\pi}  \int_{\gamma_{h,k}} M_{3}\big(e^{i(2\pi h/k+ \varphi)}\big)\
e^{-in(2\pi h/k+ \varphi)} d\varphi\ , \label{integrals} \\
&\sim  \sum_{k=1}^\infty \phi_k(n)
\end{align}
where $\gamma_{h,k}$ is an arc passing through $\varphi=0$. We don't give  a
detailed discussion on the choice of the arc but refer the interested reader to
\cite{Rademacher:1943}. In the second line, we have implicitly assumed that the
integrals and the sum over $h$  have been carried out. 

In order to carry out the integral for a particular $(h,k)$, we need to compute 
the Laurent expansion of $M_3(x)$ about the point $x=\exp(2\pi i h/k)$ and then
compute the integral using methods such as the saddle point. For usual
partitions, this is typically done using modular properties of the Dedekind eta
function. However, there is no such modular property in this case. The dominant
contribution occurs for $k=1$ (or $x=1$) and we will first consider this
contribution. 
Let
\begin{equation}
g_{3d}(t) := \log M_{3}(e^{-t})=-\tfrac12 \sum_{\nu=1}^\infty \nu(\nu+1)\ \log (1- e^{-\nu t})\equiv\sum_{\nu=1}^\infty h_{3d}(\nu) ,
\end{equation}
where $ h_{3d}(x):=  -\tfrac{x(x+1)}2\ \log (1- e^{-x t})$.The Abel-Plana
formula enables us to replace  the discrete sum over $\nu$  by the integral:
\begin{align}
g_{3d}(t) =\int_0^{\infty }h(x)\ dx -i \int_{0}^{\infty} \frac{h(iy)-h(-iy)}{e^{2\pi y} -1}\ dy\ ,
\end{align}

For $h_r(x):=-x^r\log (1- e^{-x t})$ , by expanding out the logs and resumming, 
Almkvist has shown  that\cite{Almkvist:1998}
\begin{align}
g_r(t) &=\Big[\tfrac{r!\zeta(r+2)}{t^{r+1}} + \zeta'(-r)-\zeta(-r)  \log t
+\frac{t}2 \zeta(-r-1) \Big]+ \sum_{\nu=2}^\infty
\tfrac{\zeta(1-\nu)\zeta(-r-\nu)}{\nu!}t^\nu \ , \nonumber  \\
&= \hat{g}_r(t) + g^{sum}_r(t)\ ,
\end{align}
where in the second line $g^{sum}_r(t)$ refers to terms appearing as the sum  in
the first line and $\hat{g}_r(t)$ the remaining terms (within square brackets)
up to order $t$. This separation is useful in computing the saddle-point where
we will drop the terms appearing in   $g^{sum}_r(t)$ in computing the location
of the saddle point.
Then, it follows that
\begin{equation}g_{3d}(t)=\frac12\Big(g_1(t)+g_2(t)\Big)\quad \implies \quad 
\boxed{M_3(e^{-t})\sim \exp\big[\tfrac{g_1(t)+g_2(t)}2\big]}\ .
\end{equation}
Note that the infinite sum for $ g_2(t) $ vanishes since $ \zeta(-2n)=0 $ for $
n=1,2,3,\ldots $  while for $ g_1(t) $ only terms with even $ \nu $ contribute.
In computing the integral in Eq. \eqref{integrals}, we 
\begin{equation}
\frac1{2\pi}\int_{\gamma_{1,1}}M_3\left(e^{i\varphi}\right) \ d\varphi  =
\frac{e^{\tfrac12[\zeta'(-1)+\zeta'(-2)]}  }{2\pi}
\int_{-\infty}^\infty(-i\varphi)^{-\hat{\gamma}}
e^{\left(\tfrac{a_1}{2(-i\varphi)^2} + \tfrac{2a_2}{2(-i\varphi)^3} -i\xi
\varphi\right)}\ d\varphi
\end{equation}
where $ a_1=\zeta(3) $, $ a_2=2 \zeta(4) $, $\hat{\gamma}=\zeta(-1)/2=-1/24$  
and $\xi=n+\tfrac{\zeta(-3)}4$. Using the expansion
\begin{equation}
\exp\left(\tfrac{a_1}{2(-i\varphi)^2} + \tfrac{2a_2}{2(-i\varphi)^3}\right)=
\sum_{\nu_1,\nu_2} 
\frac{a_1^{\nu_1}a_2^{\nu_2}}{2^{\nu_1+\nu_2}\nu_1!\nu_2!(-i\varphi)^{
2\nu_1+3\nu_2}}
\end{equation}
and the integral
\begin{equation}
\frac1{2\pi}\int_{-\infty}^\infty (-i\varphi)^{-\alpha} e^{-i\xi\varphi}\
d\varphi=  \left\{\begin{array}{ll}\frac{\xi^{\alpha-1}}{\Gamma(\alpha)}&
\textrm{ if } \alpha \geq 1\\ \delta(\xi) & \textrm{ if }
\alpha=0\end{array}\right.
\end{equation}
we find that the contribution ignoring the terms in $g^{sum}_{3d}(t)$  is  given by 
\begin{align}
\phi^{(0)}_1(n) & \sim \exp\big(\tfrac12[\zeta'(-1)+\zeta'(-2)]\big) 
\sum_{(\nu_1,\nu_2)\in  \mathbb{N}^2} \frac{a_1^{\nu_1}
a_2^{\nu_2}}{2^{\nu_1+\nu_2}\nu_1!\nu_2!}
\frac{\xi^{2\nu_1+3\nu_2-1+\hat{\gamma}}}{\Gamma(2\nu_1+3\nu_2+\hat{\gamma})}
\nonumber \\
&:= \exp\big(\tfrac12[\zeta'(-1)+\zeta'(-2)]\big) \  L[\xi,\hat{\gamma}]\ ,
\end{align}
where we have implicitly defined the function $L[\xi,\gamma]$ in the second
line.  In order to include the contribution of $g^{sum}_{3d}(t)$, we consider
the Taylor expansion (Note that $c_0=1$)
\begin{equation}
\exp\Big(g^{sum}_{3d}(t)\Big) = \sum_{j=0}^\infty c_j\ t^j\ ,
\end{equation} 
and carry out the integrations to obtain
\begin{align}
\phi_1(n) &= \sum_{j=0}^\infty \phi^{(j)}_1(n) \nonumber \\
&:=  \exp\big(\tfrac12[\zeta'(-1)+\zeta'(-2)]\big) \sum_{j=0}^\infty c_j\  L\left[n+\tfrac{\zeta(-3)}4,\hat{\gamma}-j\right]
\end{align}

\subsection{Other poles}

Let us evaluate $ M_{3d}\big(e^{iy}\big) $ in the neighbourhood of such a point.
 Put $ y=2\pi h/k+ \varphi $ and using a method due to Almkvist(see Theorem 5.1
in \cite{Almkvist:1998}), we get
\begin{multline}
M_{3d}\big(e^{i2\pi h/k-i\varphi}\big) \sim
\exp\Big(\tfrac12\left[\tfrac{a_1}{k^3} 
(-i\varphi)^{-2}+\tfrac{a_2}{k^4}(-i\varphi)^{-3}\right]
+\tfrac12\big[k\zeta'(-1) + k^2\zeta'(-2)\big] \\
+ \tfrac{\pi i}{2}\big[s(1,h,k)+s(2,h,k)\big] -\frac{k}2\zeta(-1)\log
(-ik\varphi)-\tfrac14\zeta(-3) i\varphi +\cdots \Big) \ ,
\end{multline}
where the generalized Dedekind sums are
\begin{align}
s(1,h,k)&= \frac{k}{3} \sum_{j=1}^{k-1} B_{2}(j/k) \log|2\sin(jh\pi/k)|  +
\frac{ik^2t}8 \sum_{j=1}^{k-1} B_3(j/k)\cot(jh\pi/k) \nonumber \\
s(2,h,k)&=\frac{k}{3} \sum_{j=1}^{k-1} B_{3}(j/k) ((jh/k))= -\frac1{16k}
\sum_{j=1}^{k-1} \cot^{(r)}(jh\pi/k) \cot(j\pi/k)\ ,
\end{align}
where $B_n(x)$ are the Bernoulli polynomials and 
\begin{equation}
((x))=\left\{ \begin{array}{lr} x-[x]-\tfrac12 & \textrm{ if }x\notin \mathbb{Z} \\
                                                0 &\textrm{ if } x \in \mathbb{Z}\end{array}\right.
\end{equation}

We illustrate the computation of $\phi_1(n)$ for $n=60$. Below, we quote the
result after rounding off  to the nearest integer and underline the number of
correct digits.
\begin{align*}
              \phi_1^{(0)}(60)&=\underline{11031}748252850258\\
\phi^{(0)}_1(60)+\phi^{(1)}_1(60)&=\underline{1103128}7052778130 \\
\phi_1(60)&=\underline{110312866}33959406\\
\phi_1(60)+\phi_2(60)&=\underline{11031286641}929870\\
       m_3(60)&=11031286641714044 
 \end{align*}
We observe that $\phi_1^{(0)}(60)$ gets the first five digits right while 
$\phi_1(60)$ makes the  estimate correct to nine digits while adding
$\phi_2(60)$ gets 11 digits right. We need to include the contributions of of
other zeros i.e., $\phi_k(n)$ for $k>2$ to further improve the estimate. We
anticipate that addition of other terms should eventually lead to an exact
answer though we have not explicitly verified that it is so.

\section{Exact enumeration of higher-dim. partitions}\label{exactresults}

In this appendix, we provide the results obtained from our exact enumeration of
three,  four and five-dimensional partitions. In all cases, we have gone
significantly beyond what is known and we have contributed our results to the
Online Encyclopedia of Integer Sequences(OEIS) -- the precise sequence is listed
in the table. We believe that it will be significantly harder to add to the
numbers of solid partitions as the generation of the last set of numbers took
around five months. In this case, adding a single number roughly doubles the
runtime. There is, however, some scope for improvement for the four and
five-dimensional partitions as the numbers were generated without
parallelization.

\begin{table}[ht]
$$
\begin{array}{|rr|rr|rr|}\hline
n & p_3(n) &n & p_3(n) & n & p_3(n)  \\[2pt] \hline
 0 & 1 & 23 & 19295226 & 46 & 8683676638832 \\
 1 & 1 & 24 & 35713454 & 47 & 14665233966068 \\
 2 & 4 & 25 & 65715094 & 48 & 24700752691832 \\
 3 & 10 & 26 & 120256653 & 49 & 41495176877972 \\
 4 & 26 & 27 & 218893580 & 50 & 69531305679518 \\
 5 & 59 & 28 & 396418699 & 51 & 116221415325837 \\
 6 & 140 & 29 & 714399381 & 52 & 193794476658112 \\
 7 & 307 & 30 & 1281403841 & 53 & 322382365507746 \\
 8 & 684 & 31 & 2287986987 & 54 & 535056771014674 \\
 9 & 1464 & 32 & 4067428375 & 55 & 886033384475166 \\
 10 & 3122 & 33 & 7200210523 & 56 & 1464009339299229 \\
 11 & 6500 & 34 & 12693890803 & 57 & 2413804282801444 \\
 12 & 13426 & 35 & 22290727268 & 58 & 3971409682633930 \\
 13 & 27248 & 36 & 38993410516 & 59 & 6520649543912193 \\
 14 & 54804 & 37 & 67959010130 & 60 & 10684614225715559 \\
 15 & 108802 & 38 & 118016656268 & 61 & 17472947006257293 \\
 16 & 214071 & 39 & 204233654229 & 62 & 28518691093388854 \\
 17 & 416849 & 40 & 352245710866 & 63 & 46458506464748807 \\
 18 & 805124 & 41 & 605538866862 & 64 & 75542021868032878 \\
 19 & 1541637 & 42 & 1037668522922 & 65 & 122606799866017598 \\
 20 & 2930329 & 43 & 1772700955975 & 66 & 198635761249922839 \\
 21 & 5528733 & 44 & 3019333854177 & 67 & 321241075686259326 \\
 22 & 10362312 & 45 & 5127694484375 & 68 & 518619444932991189 \\ \hline
\end{array}
$$
\caption{Numbers of solid partitions. This is sequence A000293 in the OEIS\cite{OEIS}.}
\end{table}
\begin{table}[ht]
$$
\begin{array}{|rr|rr|rr|}\hline
n & p_4(n) &n & p_4(n) & n & p_4(n)  \\[2pt] \hline
 0 & 1 & 13 & 181975 & 25 & 2569270050 \\
 1 & 1 & 14 & 425490 & 26 & 5427963902 \\
 2 & 5 & 15 & 982615 & 27 & 11404408525 \\
 3 & 15 & 16 & 2245444 & 28 & 23836421895 \\
 4 & 45 & 17 & 5077090 & 29 & 49573316740 \\
 5 & 120 & 18 & 11371250 & 30 & 102610460240 \\
 6 & 326 & 19 & 25235790 & 31 & 211425606778 \\
 7 & 835 & 20 & 55536870 & 32 & 433734343316 \\
 8 & 2145 & 21 & 121250185 & 33 & 886051842960 \\
 9 & 5345 & 22 & 262769080 & 34 & 1802710594415 \\
 10 & 13220 & 23 & 565502405 & 35 & 3653256942840 \\
 11 & 32068 & 24 & 1209096875 &  &  \\
 12 & 76965 & 25 & 2569270050 &  & \\ \hline
\end{array}$$
\caption{Numbers of four-dimensional  partitions. This is sequence A000334 in the OEIS\cite{OEIS}.}
\end{table}
 \begin{table}[ht]
$$
\begin{array}{|rr|rr|rr|}\hline
n & p_5(n) &n & p_5(n) & n & p_5(n)  \\[2pt] \hline
 0 & 1 & 11 & 119140 & 22 & 3923114261 \\
 1 & 1 & 12 & 323946 & 23 & 9554122089 \\
 2 & 6 & 13 & 869476 & 24 & 23098084695 \\
 3 & 21 & 14 & 2308071 & 25 & 55458417125 \\
 4 & 71 & 15 & 6056581 & 26 & 132293945737 \\
 5 & 216 & 16 & 15724170 & 27 & 313657570114 \\
 6 & 657 & 17 & 40393693 & 28 & 739380021561 \\
 7 & 1907 & 18 & 102736274 & 29 & 1733472734334 \\
 8 & 5507 & 19 & 258790004 & 30 & 4043288324470 \\
 9 & 15522 & 20 & 645968054 &  &  \\
 10 & 43352 & 21 & 1598460229 &  &  \\ \hline
\end{array}
$$
\caption{Numbers of five-dimensional  partitions. This is sequence A000390 in the OEIS\cite{OEIS}.}
\end{table}
\clearpage
\bibliography{partitions}

\providecommand{\href}[2]{#2}\begingroup\raggedright\begin{thebibliography}{10}

\bibitem{Hardy:1917}
G.~H. Hardy and S.~Ramanujan, ``Asymptotic formul\ae\ in combinatory analysis
  [{P}roc. {L}ondon {M}ath. {S}oc. (2) {\bf 16} (1917),,'' in {\em Collected
  papers of {S}rinivasa {R}amanujan}, p.~244.
\newblock AMS Chelsea Publ., Providence, RI, 2000.

\bibitem{Rademacher:1937}
H.~Rademacher, ``On the partition function {$p(n)$},'' {\em Proc. London Math.
  Soc.} {\bfseries 43} (1937) 241--254.

\bibitem{MacMahon}
P.~A. MacMahon, {\em Combinatory analysis. {V}ol. {I}, {II} (bound in one
  volume)}.
\newblock Dover Phoenix Editions. Dover Publications Inc., Mineola, NY, 2004.
\newblock Reprint of {{\it An introduction to combinatory analysis}} (1920) and
  {{\it Combinatory analysis. Vol. I, II}} (1915, 1916).

\bibitem{Wright:1931}
E.~M. Wright, ``Asymptotic partition formulae. I, Plane partitions,'' {\em
  Quart. J. Math.} {\bfseries 2} (1931) 177--189.

\bibitem{Atkin:1967}
A.~O.~L. Atkin, P.~Bratley, I.~G. Macdonald, and J.~K.~S. McKay, ``Some
  computations for {$m$}-dimensional partitions,'' {\em Proc. Cambridge Philos.
  Soc.} {\bfseries 63} (1967) 1097--1100.

\bibitem{Wu:1997}
F.~Y. Wu, ``The infinite-state Potts model and restricted multidimensional
  partitions of an integer,''
{\em Math. Comput. Modelling} {\bfseries 26} (1997) 269Ñ274.

\bibitem{Huang:1997}
H.~Y. Huang and F.~Y. Wu,
  \href{http://dx.doi.org/10.1142/S0217979297000150}{``The infinite-state
  {P}otts model and solid partitions of an integer,''} in {\em Proceedings of
  the {C}onference on {E}xactly {S}oluble {M}odels in {S}tatistical
  {M}echanics: {H}istorical {P}erspectives and {C}urrent {S}tatus ({B}oston,
  {MA}, 1996)}, vol.~11, pp.~121--126.
\newblock 1997.

\bibitem{Bhatia:1997}
D.~P. Bhatia, M.~A. Prasad, and D.~Arora, ``Asymptotic results for the number
  of multidimensional partitions of an integer and directed compact lattice
  animals,'' \href{http://dx.doi.org/10.1088/0305-4470/30/7/010}{{\em J. Phys.
  A} {\bfseries 30} no.~7, (1997) 2281--2285}.

\bibitem{Feng:2007ur}
B.~Feng, A.~Hanany, and Y.-H. He, ``{Counting gauge invariants: The Plethystic
  program},'' \href{http://dx.doi.org/10.1088/1126-6708/2007/03/090}{{\em JHEP}
  {\bfseries 0703} (2007) 090},
  \href{http://arxiv.org/abs/hep-th/0701063}{{\ttfamily arXiv:hep-th/0701063
  [hep-th]}}.

\bibitem{Lucietti:2008cv}
J.~Lucietti and M.~Rangamani, ``{Asymptotic counting of BPS operators in
  superconformal field theories},''
  \href{http://dx.doi.org/10.1063/1.2970775}{{\em J.Math.Phys.} {\bfseries 49}
  (2008) 082301}, \href{http://arxiv.org/abs/0802.3015}{{\ttfamily
  arXiv:0802.3015 [hep-th]}}.

\bibitem{Gopakumar:1998ii}
R.~Gopakumar and C.~Vafa, ``{M theory and topological strings. 1.},''
  \href{http://arxiv.org/abs/hep-th/9809187}{{\ttfamily arXiv:hep-th/9809187
  [hep-th]}}.

\bibitem{Gopakumar:1998jq}
R.~Gopakumar and C.~Vafa, ``{M theory and topological strings. 2.},''
  \href{http://arxiv.org/abs/hep-th/9812127}{{\ttfamily arXiv:hep-th/9812127
  [hep-th]}}.

\bibitem{Behrend:2009}
K.~Behrend, J.~Bryan, and B.~Szendroi, ``Motivic degree zero Donaldson-Thomas
  invariants,''
\href{http://arxiv.org/abs/0909.5088}{{\ttfamily arXiv:0909.5088 [math-ag]}}.

\bibitem{Mustonen:2003}
V.~Mustonen and R.~Rajesh, ``Numerical estimation of the asymptotic behaviour
  of solid partitions of an integer,''
  \href{http://dx.doi.org/10.1088/0305-4470/36/24/304}{{\em J. Phys. A}
  {\bfseries 36} no.~24, (2003) 6651--6659}.

\bibitem{AndrewsPartitions}
G.~E. Andrews, {\em The theory of partitions}.
\newblock Cambridge Mathematical Library. Cambridge University Press,
  Cambridge, 1998.
\newblock Reprint of the 1976 original.

\bibitem{Stanley:1971}
R.~P. Stanley, ``Theory and application of plane partitions. {I}, {II},'' {\em
  Studies in Appl. Math.} {\bfseries 50} (1971) 167--188; ibid. 50 (1971),
  259--279.

\bibitem{Stanley:1985}
R.~P. Stanley, ``Plane partitions: past, present, and future,'' in {\em
  Combinatorial {M}athematics: {P}roceedings of the {T}hird {I}nternational
  {C}onference ({N}ew {Y}ork, 1985)}, vol.~555 of {\em Ann. New York Acad.
  Sci.}, pp.~397--401.
\newblock New York Acad. Sci., New York, 1989.

\bibitem{Wilf:2000}
H.~S. Wilf, ``Lectures on integer partitions,'' tech. rep., University of
  Pennsylvania, 2000.
\newblock Lectures by H.S. Wilf at the U. of Victoria in 2000 available at
  http://cis.upenn.edu/$\widetilde{\mbox{~}}$wilf.

\bibitem{Finch:2004}
S.~Finch, ``Integer Partitions,'' tech. rep.,
  http://algo.inria.fr/csolve/prt.pdf, 2004.

\bibitem{Knuth:1970}
D.~E. Knuth, ``A note on solid partitions,'' {\em Math. Comp.} {\bfseries 24}
  (1970) 955--961.

\bibitem{Almkvist:1993}
G.~Almkvist, ``A rather exact formula for the number of plane partitions,'' in
  {\em A tribute to {E}mil {G}rosswald: number theory and related analysis},
  vol.~143 of {\em Contemp. Math.}, pp.~21--26.
\newblock Amer. Math. Soc., Providence, RI, 1993.

\bibitem{Almkvist:1998}
G.~Almkvist, ``Asymptotic formulas and generalized {D}edekind sums,'' {\em
  Experiment. Math.} {\bfseries 7} no.~4, (1998) 343--359.

\bibitem{GG}
D.~S. Gaunt and A.~J. Guttmann, ``Asymptotic analysis of coefficients,'' in
  {\em Phase transitions and critical points}, C.~Domb and M.~S. Green, eds.,
  vol.~3, ch.~4, pp.~181--243.
\newblock Academic Press, New York, 1974.

\bibitem{Widom:2002}
M.~Widom, R.~Mosseri, N.~Destainville, and F.~Bailly, ``Arctic Octahedron in
  Three-Dimensional Rhombus Tilings and Related Integer Solid Partitions,''
  {\em J. of Stat. Phys.} {\bfseries 109} no.~516, (2002) 945--965.

\bibitem{Bratley:1967a}
P.~Bratley and J.~K.~S. McKay, ``Algorithm 313: Multi-dimensional partition
  generator.,'' {\em Commun. ACM} (1967) 1--1.

\bibitem{Stanley:1986}
R.~P. Stanley, ``Symmetries of plane partitions,''
  \href{http://dx.doi.org/10.1016/0097-3165(86)90028-2}{{\em J. Combin. Theory
  Ser. A} {\bfseries 43} no.~1, (1986) 103--113}. Erratum: ibid. 44 (1987), no.
  2, 310.

\bibitem{Krattenthaler:1990}
C.~Krattenthaler, ``Generating functions for plane partitions of a given
  shape,'' \href{http://dx.doi.org/10.1007/BF02567918}{{\em Manuscripta Math.}
  {\bfseries 69} no.~2, (1990) 173--201}.

\bibitem{Bender:1972}
E.~A. Bender and D.~E. Knuth, ``Enumeration of plane partitions,'' {\em J.
  Combinatorial Theory Ser. A} {\bfseries 13} (1972) 40--54.

\bibitem{Nijenhuis:1978}
A.~Nijenhuis and H.~S. Wilf, {\em Combinatorial algorithms}.
\newblock Academic Press Inc. [Harcourt Brace Jovanovich Publishers], New York,
  second~ed., 1978.
\newblock For computers and calculators, Computer Science and Applied
  Mathematics.

\bibitem{Meinardus:1954}
G.~Meinardus, ``Asymptotische {A}ussagen \"uber {P}artitionen,'' {\em Math. Z.}
  {\bfseries 59} (1954) 388--398.

\bibitem{Rademacher:1943}
H.~Rademacher, ``On the expansion of the partition function in a series,'' {\em
  Ann. of Math. (2)} {\bfseries 44} (1943) 416--422.

\bibitem{OEIS}
``The On-line Encyclopedia of Integer Sequences,'' 2011.
\newblock published electronically at http://oeis.org.

\end{thebibliography}\endgroup
\end{document}